\documentclass[conference]{IEEEtran}
\IEEEoverridecommandlockouts
\usepackage{cite}
\usepackage{amsmath,amssymb,amsfonts}
\usepackage{algorithmic}
\usepackage{graphicx}
\usepackage{multirow}
\usepackage{textcomp}
\usepackage[dvipsnames]{xcolor}
\definecolor{myyellow}{HTML}{FFC000}
\definecolor{mygreen}{HTML}{00B050}
\definecolor{myblue}{HTML}{0070C0}
\usepackage{url}

\def\BibTeX{{\rm B\kern-.05em{\sc i\kern-.025em b}\kern-.08em
    T\kern-.1667em\lower.7ex\hbox{E}\kern-.125emX}}
\begin{document}

\title{The Beauty or the Beast: Which Aspect of Synthetic Medical Images Deserves Our Focus?\
}



\author{\IEEEauthorblockN{1\textsuperscript{st} Xiaodan Xing}
\IEEEauthorblockA{\textit{National Heart and Lung Institute} \\
\textit{Imperial College London}\\
London, U.K. \\
x.xing@imperial.ac.uk}
\and
\IEEEauthorblockN{2\textsuperscript{nd} Yang Nan}
\IEEEauthorblockA{\textit{National Heart and Lung Institute} \\
\textit{Imperial College London}\\
London, U.K. \\
y.nan20@imperial.ac.uk}
\and
\IEEEauthorblockN{3\textsuperscript{rd} Federico Felder}
\IEEEauthorblockA{\textit{National Heart and Lung Institute} \\
\textit{Imperial College London}\\
London, U.K. \\
f.felder@imperial.ac.uk}
\and
\IEEEauthorblockN{4\textsuperscript{th} Simon Walsh}
\IEEEauthorblockA{\textit{National Heart and Lung Institute} \\
\textit{Imperial College London}\\
London, U.K. \\
s.walsh@imperial.ac.uk}
\and
\IEEEauthorblockN{5\textsuperscript{th} Guang Yang}
\IEEEauthorblockA{\textit{National Heart and Lung Institute} \\
\textit{Imperial College London}\\
London, U.K. \\
g.yang@imperial.ac.uk}

}

\maketitle

\begin{abstract}

Training medical AI algorithms requires large volumes of accurately labeled datasets, which are difficult to obtain in the real world. Synthetic images generated from deep generative models can help alleviate the data scarcity problem, but their effectiveness relies on their fidelity to real-world images. Typically, researchers select synthesis models based on image quality measurements, prioritizing synthetic images that appear realistic. However, our empirical analysis shows that high-fidelity and visually appealing synthetic images are not necessarily superior. In fact, we present a case where low-fidelity synthetic images outperformed their high-fidelity counterparts in downstream tasks. Our findings highlight the importance of comprehensive analysis before incorporating synthetic data into real-world applications. We hope our results will raise awareness among the research community of the value of low-fidelity synthetic images in medical AI algorithm training.

\end{abstract}


\section{Introduction}
Synthetic data can solve data scarcity problem by generating more samples for the training dataset. Commonly, deep learning practitioners favor synthetic models with better visualization performance, while this evaluation can be subjective and non-reproducible. Pre-defined metrics evaluating the synthesis performance were then proposed. 

Metrics evaluating the synthesis performance can vary. Most metrics evaluate the fidelity of synthetic images, i.e., whether the distribution of synthetic images is similar to real distributions. Fréchet Inception Distance (FID) \cite{heusel2017gans} and Inception Score (IS) \cite{salimans2016improved} are the two most practiced fidelity measurements. Besides, precision, recall, and F1-score \cite{lucic2018gans} were also implemented in image synthesis scenarios to further provide a gauge for the variety of synthetic images. A high recall value indicates that a synthetic model can generate all patterns from the real dataset, i.e., capturing a wide variety of real datasets. All these measurements have been advocated for consistency with human perception, and selecting synthetic models based on these metrics has then been commonly practiced.

Synthetic images with high fidelity and variety scores can intuitively be used to indirectly measure the utility of synthetic images \cite{tran2021data}, while the reality is often more complex than a simple correlation. 

In this study, we compared three state-of-the-art deep generative models and analyzed their quality using FID, precision, and recall values. To evaluate the utility of synthetic images, two strategies were used in this work: (1) the data augmentation utility — we added the synthetic data to the training dataset and observed the classification improvement brought by additional synthetic data; and (2) the feature extraction utility — we pre-trained classification models on fully synthetic datasets and fine-tuned the last layer of these classification models on a small scale of real datasets. This strategy measures whether synthetic images can produce powerful features that facilitate downstream tasks. 

From experiments on two widely used public medical image datasets, we empirically show that the fidelity measurement does not correlate with utility as previously assumed. We provide an example of how high fidelity images failed to contribute to generalizable data augmentation performance. 

\section{Deep Generative Models and Evaluation Metrics}
\begin{figure}[h]
\includegraphics[width=\columnwidth]{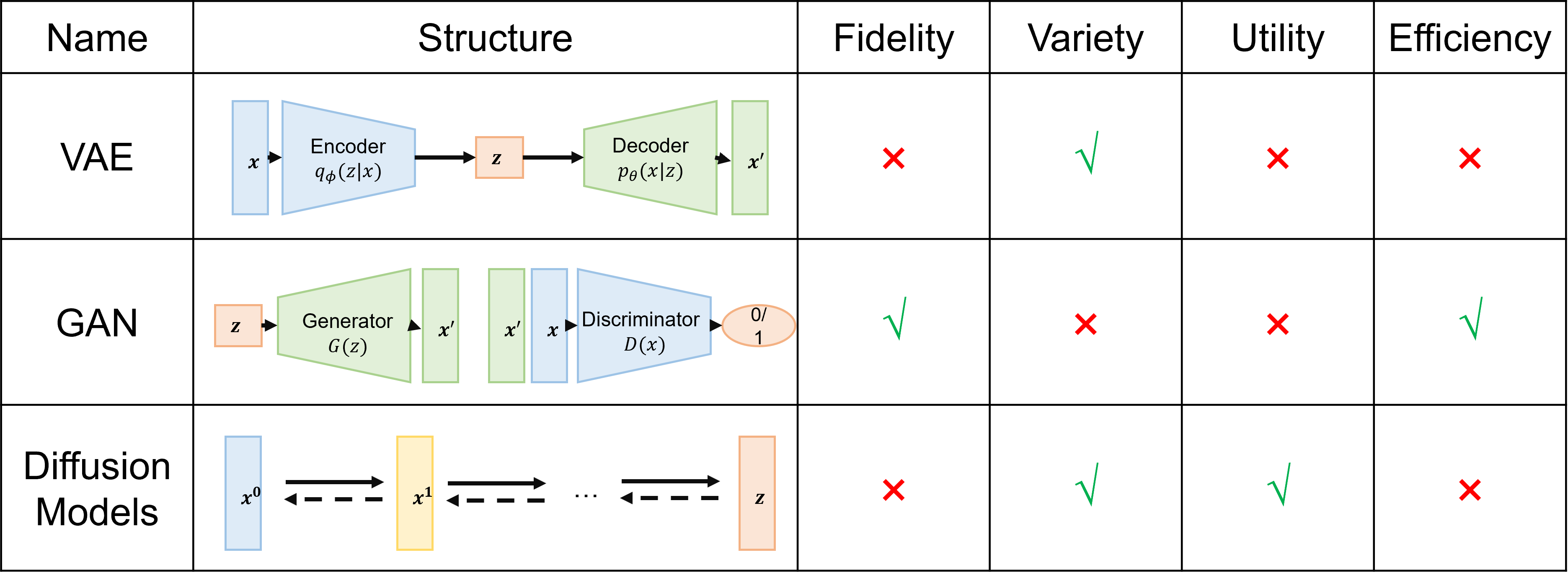}
\caption{Deep generative models validated in this study and their performances.} \label{fig:teaser}
\end{figure}

This study focuses on the development and evaluation of the fidelity and utility of three state-of-the-art deep generative models. Figure \ref{fig:teaser} shows the basic architecture of these models, and we present our experimental results regarding their performances.
\subsection{Variantional Auto-Encoders (VAE)}
An auto-encoder is composed of two parts: an encoder that maps the input images into a lower dimensional latent feature space and a decoder that reconstructs the images from the latent space. By sampling from the latent space, one could generate new images from the real data distributions. VAEs simplify the sampling procedure by assuming that the latent space follows prior distributions. Compared to GAN models, VAEs explicitly model the latent distributions with parametric distributions, such as Gaussian distributions, and increase the interpretability and controllability of generative models. An additional bonus of this explicit modeling is that VAEs can generate samples with higher variances compared to GAN models and do not often suffer from mode collapses \cite{kingma2013auto}. 

In this study, we chose Vector-Quantized VAE (VQ-VAE) \cite{van2017neural,razavi2019generating} as a representative method for the VAE family and implemented VQ-VAE2 to generate synthetic images. VQ-VAE can solve the blurring problem during reconstructions and address the distribution approximation problem with autoregressive models. Instead of using parametric distributions to model the latent space, VQ-VAE approximates the latent space distributions with pre-trained networks. However, finding a suitable prior distribution for continuous variables can be complicated, and the joint or conditional distributions among multiple continuous variables are difficult to derive from data-driven methods. Thus, VQ-VAE quantizes the latent features into a discrete latent space, i.e., each pixel in the latent feature maps is a $K$-way categorical variable, sampling from 0 to $K$, and by using autoregressive models that compute the conditional distributions among pixels, the latent space distribution could be approximated. 
\subsection{GAN-Based Models}
GAN-based models have been the most popular backbone for image synthesis since 2014 \cite{goodfellow2020generative}. Featured by two networks training in an adversarial way, GAN and its variants have been proven to be efficient in high-resolution medical image synthesis. In this study, we implemented StyleGAN2 as a representative model for the GAN family because StyleGAN2 methods have brought new standards for generative modeling regarding image quality \cite{sauer2022stylegan}.


However, GAN-based models are cursed by the mode collapse problem, and both StyleGAN1 \cite{karras2019style} and StyleGAN2 \cite{karras2020training} are still hobbled by this issue. The vanilla discriminator loss of GAN models only optimizes the fidelity of synthetic images, i.e., the similarity between real and synthetic images. It introduces multiple minimal values for the discriminator loss when the real datasets are varied. Without further constraints, it is possible for the generator to reach only one of the minima and produce similar outputs. Loss functions such as Wasserstein loss \cite{arjovsky2017wasserstein} could alleviate the mode collapse problem, but the loss of variety during synthesis cannot be fully resolved. 
\subsection{Diffusion Models}
To balance the variety and performance of deep synthesis models, diffusion models \cite{rombach2022high} have been proposed and have shown great potential in various high-quality image syntheses. Diffusion models gradually downgrade real images with Gaussian noises and use neural networks to recover real images from downgraded images. By doing so, the neural networks obtain the ability to recover real images from noises. However, the sequential evaluation process of the diffusion models requires hundreds of GPU days to optimize and consume large-scale computation resources. 

To reduce the time and memory costs of training, latent diffusion models (LDMs) \cite{rombach2022high} have been proposed. The LDM operates the diffusion process in the latent space and enables the optimization of diffusion models on limited computational resources. In this study, we implemented LDM with VQ-VAE for latent space feature extraction. 
\subsection{Fidelity, Variety, and Utility}
In this study, we evaluated the performance of synthetic images in three aspects. The first was image fidelity, i.e., the similarity between real images and synthetic images. We invited two clinical experts with different years of experience to discriminate synthetic images from real images and summarized their discrimination accuracy as one of the fidelity scores. In addition, we used the FID score to measure the distributions between real and synthetic images. 

To assess the variety of synthetic images, we used the JPEG file size of the mean image and the precision-recall metrics. The lossless JPEG file size of the group average image was used to measure the inner class variety in the ImageNet dataset \cite{deng2009imagenet}. This operation was justified by presuming that a dataset containing diverse images would result in a blurrier average image, and therefore, would reduce the loss-less JPEG file size of the mean image.

We also implemented precision and recall metrics to statistically compute the diversity of synthetic images. Essentially, the precision of a synthetic model was the ratio of realistic synthetic images to all synthetic images, and the recall of a synthetic model corresponded to the ratio of real images whose mode was covered by synthetic images to all real images. Precision and recall reflected fidelity and variety, respectively. Both the file size of mean images and the recall measured the “variety”, while they focused on different aspects of the dataset diversity. The file size of mean images measured the variety among synthetic datasets and favored models that produced various images, while the recall favored models that produced synthetic images whose variety was similar to the real data distribution. 

For the utility, we focused on data augmentation and feature extraction utilities to simulate real-world use cases of synthetic data. The data augmentation utility was measured by the improved classification accuracy when adding additional synthetic data into the training dataset. A paired Wilcoxon Signed Rank Test was performed to evaluate the significance of accuracy improvement. According to this evaluation, useful synthetic images were supposed to bring significant accuracy improvements. 

The feature extraction utility was measured by the classification accuracies of models pre-trained on synthetic datasets and fine-tuned on real datasets. During the fine-tuning, we froze the gradients of all layers except for the last fully connected layer. By doing so, we evaluated whether synthetic images could provide powerful features that facilitate downstream tasks. In the feature extraction evaluation, useful synthetic images were assumed to produce accurate classification results and had no significant difference compared to the models trained on real datasets. 

\section{Experimental Settings and Parameters}
\textbf{Dataset and Pre-Processing.} We evaluated the synthetic performance of two datasets containing both RGB and greyscale medical images. The first dataset contained 252 hematoxylin-eosin (H\&E) stained whole-slide images (WSIs) from the breast cancer semantic segmentation (BCSS) database \cite{amgad2019structured} and the Lizard database \cite{graham2021lizard}. WSIs were cropped into small patches with a size of $256\times256$, and we further classified these patches into 6 categories including inflammatory, necrosis, stroma, tumor, fat, and gland. Overall, 18,703 patches were obtained, and we split the patches into training (7,874 patches), validation (3,741 patches), and independent testing (7,478 patches) subsets. 

The second dataset was an X-ray dataset\footnote{https://www.kaggle.com/datasets/paultimothymooney/chest-xray-pneumonia?resource=download} containing 5,863 chest X-ray images of pneumonia patients and normal controls. All X-ray images were obtained in the anterior-posterior order and from pediatric patients one to five years old from Guangzhou Women and Children’s Medical Center, Guangzhou. The clinical decisions were made and re-checked twice by expert physicians. The original resolution of X-ray images ranged from $[127, 2916]$. However, we resized the X-ray images to a resolution of $512\times512$. We employed the dataset division strategy in the original dataset to assure a compatible classification performance compared with other algorithms. Overall, 5,216 images were used for training, 16 were used for validation, and 624 for independent testing. To avoid information leakage, we only used training subsets for synthetic image generation. 

To further compare the synthesis performance between StyleGAN2 and LDM, we computed the quality metrics of synthetic images on an additional CT dataset \cite{walsh2018deep}. We used CT montages to represent the 3D CT images. For each scan, the top and bottom slices that had lung regions $\leq$ 10\% were discarded because of limited tissue information. We then divided the remaining slices into 4 equal clusters according to their axial position. For each cluster, one slice was randomly selected, and we tiled the 4 selected slices from all clusters into a 4-image 2D CT montage. For each 3D volume, 20 montages were generated, and overall, we obtained 52,540 montages, from which we selected 26,270 for the training subsets and 26,270 for the quality evaluation of deep generative models. 

\textbf{Implementation Details. }In this study, we implemented three architectures for medical image synthesis, including VQ-VAE, StyleGAN2, and LDM. The implementation codes will be publicized in \url{https://github.com/XiaodanXing/CBMS2023_synthetic_data}. 

For the VQ-VAE2 models, we used a two-level latent hierarchy with feature maps of size $32^2$ and $64^2$. Since conditional implementation of VQ-VAE2 would increase the inference time and decrease the synthesis performance, thus we trained VQ-VAE2 models for each image category respectively. For the LDMs on $256^2$ and $512^2$, we first used a VQ-VAE encoder to compress the images by 8 times. For both StyleGAN2 and LDM, we used a conditional training strategy. The image categories were encoded into a latent vector of 512. 

\noindent\textbf{Evaluation Metric Computation.} The first fidelity score is named as fake identification rate (FIR). This human quality measurement was performed on small sub-subsets. We selected 2 images (pathological dataset) and 5 cases (X-ray dataset) for each category and 10 cases (CT dataset) from each synthetic algorithm and the training dataset and augmented training dataset, resulting in 60 pathological image, 50 X-ray images and 50 CT images. Then, we shuffle the image order and invited two human experts (one clinician with five years of experience and one technician with one year of experience) to identify synthetic images on these two sub-subsets. We allow the human experts to discuss their opinions and produced one result (fake/real) for each image. 

The FID was computed on features for the last fully connected layer extracted from pre-trained InceptionV3 on the ImageNet dataset, and the sizes of the extracted features are 2048×1. We also used the extracted features to compute the precision and recall values.

For the utility measurement, classification networks based on InceptionV3 were trained on the training subsets, and we selected the best performed models on the validation subsets. We compared the augmentation utility with a combination of traditional augmentation methods, including random flipping, rotating, and contrast changing. For the feature extraction utility, we selected 50\% of the training dataset to fine-tune the models pre-trained on synthetic subsets, and we saved the classification models after 20 epochs. 
\section{Experimental Results}
\subsection{Comparing Qualities of Different Synthesis Models }	
We presented the evaluation results on the pathological dataset (Table \ref{tab:pathology}) and the X-ray dataset (Table \ref{tab:xray}), respectively. All quality measurements were computed between the distributions of generated images and the distributions of real images from the testing dataset. We also computed the quality measurements between the training dataset and the testing dataset as a reference. Poor performance such as \textcolor{red}{low fidelity}, \textcolor{myyellow}{low variety}, \textcolor{mygreen}{low utility}, and \textcolor{myblue}{low efficiency} were highlighted with corresponding colors. * indicates a p-value $<$0.05 compared to the model trained only on the training dataset (the first rows of Table \ref{tab:pathology},\ref{tab:xray}, \ref{tab:xraycropped}).

\begin{table}[h]
\caption{Fidelity, variety, utility, and efficiency measurements of all deep generative models in the pathological dataset.}\label{tab:pathology}
\resizebox{\linewidth}{!}{\begin{tabular}{|l|lll|ll|ll|ll|}
\hline
 & \multicolumn{3}{l|}{Fidelity} & \multicolumn{2}{l|}{Variety} & \multicolumn{2}{l|}{Utility} & \multicolumn{2}{l|}{Efficiency} \\ \cline{2-10} 
\multirow{-2}{*}{Method} & \multicolumn{1}{l|}{FIR} & \multicolumn{1}{l|}{FID} & Precision & \multicolumn{1}{l|}{Recall} & File size & \multicolumn{1}{l|}{\begin{tabular}[c]{@{}l@{}}Augment \\ (\%)\end{tabular}} & \begin{tabular}[c]{@{}l@{}}Extraction\\  (\%)\end{tabular} & \multicolumn{1}{l|}{\begin{tabular}[c]{@{}l@{}}GB memory  \\ (8 images)\end{tabular}} & \begin{tabular}[c]{@{}l@{}}Inference time \\ (1000 images)\end{tabular} \\ \hline
Training & \multicolumn{1}{l|}{0.00} & \multicolumn{1}{l|}{22.11} & 0.65 & \multicolumn{1}{l|}{0.66} & 38.53 & \multicolumn{1}{l|}{84.22} & 84.22 & \multicolumn{1}{l|}{/} & / \\ \hline
Augmented & \multicolumn{1}{l|}{0.17} & \multicolumn{1}{l|}{{\color[HTML]{FF0000} 185.89}} & {\color[HTML]{FF0000} 0.04} & \multicolumn{1}{l|}{0.71} & 48.93 & \multicolumn{1}{l|}{(+) 4.07*} & / & \multicolumn{1}{l|}{/} & / \\ \hline
VQ-VAE2 & \multicolumn{1}{l|}{{\color[HTML]{FF0000} 0.83}} & \multicolumn{1}{l|}{{\color[HTML]{FF0000} 201.85}} & 0.51 & \multicolumn{1}{l|}{{\color[HTML]{FFC000} 0.16}} & 44.29 & \multicolumn{1}{l|}{{\color[HTML]{00B050} (-) 5.73}} & {\color[HTML]{00B050} (-) 21.10*} & \multicolumn{1}{l|}{{\color[HTML]{0070C0} 12.25}} & {\color[HTML]{0070C0} 12 h 6 min} \\ \hline
StyleGAN2 & \multicolumn{1}{l|}{0.08} & \multicolumn{1}{l|}{82.38} & 0.57 & \multicolumn{1}{l|}{{\color[HTML]{FFC000} 0.21}} & {\color[HTML]{FFC000} 114.67} & \multicolumn{1}{l|}{{\color[HTML]{00B050} (+) 0.08}} & {\color[HTML]{00B050} (-) 19.77*} & \multicolumn{1}{l|}{0.87} & 1   min \\ \hline
LDM & \multicolumn{1}{l|}{0.25} & \multicolumn{1}{l|}{62.56} & 0.46 & \multicolumn{1}{l|}{0.43} & 44.09 & \multicolumn{1}{l|}{(+) 3.64*} & {\color[HTML]{00B050} (-) 3.19*} & \multicolumn{1}{l|}{{\color[HTML]{0070C0} 18.1}} & {\color[HTML]{0070C0} 12   min} \\ \hline
\end{tabular}}

\end{table}

\begin{table}[h]
\caption{Fidelity, variety, utility, and efficiency measurements of all deep generative models in the X-ray dataset. }\label{tab:xray}
\resizebox{\linewidth}{!}{\begin{tabular}{|l|lll|ll|ll|ll|}
\hline
 & \multicolumn{3}{l|}{Fidelity} & \multicolumn{2}{l|}{Variety} & \multicolumn{2}{l|}{Utility} & \multicolumn{2}{l|}{Efficiency} \\ \cline{2-10} 
\multirow{-2}{*}{Method} &  \multicolumn{1}{l|}{FIR} & \multicolumn{1}{l|}{FID} & Precision & \multicolumn{1}{l|}{Recall} & File size & \multicolumn{1}{l|}{\begin{tabular}[c]{@{}l@{}}Augment \\ (\%)\end{tabular}} & \begin{tabular}[c]{@{}l@{}}Extraction\\  (\%)\end{tabular} & \multicolumn{1}{l|}{\begin{tabular}[c]{@{}l@{}}GB memory  \\ (8 images)\end{tabular}} & \begin{tabular}[c]{@{}l@{}}Inference time \\ (1000 images)\end{tabular} \\ \hline
Training & \multicolumn{1}{l|}{{\color[HTML]{333333} 0.40}} & \multicolumn{1}{l|}{{\color[HTML]{333333} 6.06}} & {\color[HTML]{333333} 0.73} & \multicolumn{1}{l|}{{\color[HTML]{333333} 0.8}} & {\color[HTML]{333333} 50.46} & \multicolumn{1}{l|}{{\color[HTML]{333333} 84.77\%}} & {\color[HTML]{333333} 84.77} & \multicolumn{1}{l|}{{\color[HTML]{333333} /}} & {\color[HTML]{333333} /} \\ \hline
Augmented & \multicolumn{1}{l|}{{\color[HTML]{333333} 0.50}} & \multicolumn{1}{l|}{{\color[HTML]{FF0000} 35.45}} & {\color[HTML]{FF0000} 0.03} & \multicolumn{1}{l|}{{\color[HTML]{333333} 0.8}} & {\color[HTML]{333333} 51.58} & \multicolumn{1}{l|}{{\color[HTML]{333333} (-) 5.12*}} & {\color[HTML]{333333} /} & \multicolumn{1}{l|}{{\color[HTML]{333333} /}} & {\color[HTML]{333333} /} \\ \hline
VQ-VAE2 & \multicolumn{1}{l|}{{\color[HTML]{FF0000} 1.00}} & \multicolumn{1}{l|}{{\color[HTML]{FF0000} 45.59}} & {\color[HTML]{FF0000} 0.00} & \multicolumn{1}{l|}{{\color[HTML]{FFC000} 0.00}} & {\color[HTML]{333333} 54.34} & \multicolumn{1}{l|}{{\color[HTML]{00B050} (+) 0.96}} & {\color[HTML]{00B050} (-) 22.27*} & \multicolumn{1}{l|}{{\color[HTML]{00B0F0} 22.82}} & {\color[HTML]{00B0F0} 12h 30min} \\ \hline
StyleGAN2 & \multicolumn{1}{l|}{{\color[HTML]{FF0000} 0.70}} & \multicolumn{1}{l|}{{\color[HTML]{333333} 9.85}} & {\color[HTML]{333333} 0.68} & \multicolumn{1}{l|}{{\color[HTML]{FFC000} 0.19}} & {\color[HTML]{FFC000} 115.62} & \multicolumn{1}{l|}{{\color[HTML]{00B050} (-) 1.92}} & {\color[HTML]{333333} (+) 1.76} & \multicolumn{1}{l|}{{\color[HTML]{333333} 1.19}} & {\color[HTML]{333333} 1 min} \\ \hline
LDM & \multicolumn{1}{l|}{{\color[HTML]{FF0000} 0.80}} & \multicolumn{1}{l|}{{\color[HTML]{FF0000} 25.68}} & {\color[HTML]{FF0000} 0.19} & \multicolumn{1}{l|}{{\color[HTML]{FFC000} 0.04}} & {\color[HTML]{FFC000} 70.59} & \multicolumn{1}{l|}{{\color[HTML]{00B050} (-) 0.64}} & {\color[HTML]{333333} (-) 0.80} & \multicolumn{1}{l|}{{\color[HTML]{00B0F0} 70.61}} & {\color[HTML]{00B0F0} 39 min} \\ \hline
\end{tabular}}

\end{table}
For both datasets, the VQ-VAE2 model preserved the inner variety distribution of real datasets, while it failed to capture the real dataset distribution. Ten out of twelve VQ-VAE2 synthesized images were successfully identified (0.83 FIR in Table \ref{tab:pathology}), and all of VQ-VAE2 synthesized images were successfully identfied by humans (1.00 FIR in Table \ref{tab:xray}). In this study, we implemented VQ-VAE2, and previous studies have shown that the quantized feature space and the hierarchy of feature space modeling could improve the performance of VAE \cite{razavi2019generating}, but our results showed that VQ-VAE2 still failed to demonstrate superior FID scores (201.85 in Table \ref{tab:pathology}, 45.59 in Table \ref{tab:xray}) compared to other synthesis methods. 

\begin{figure}
\includegraphics[width=\columnwidth]{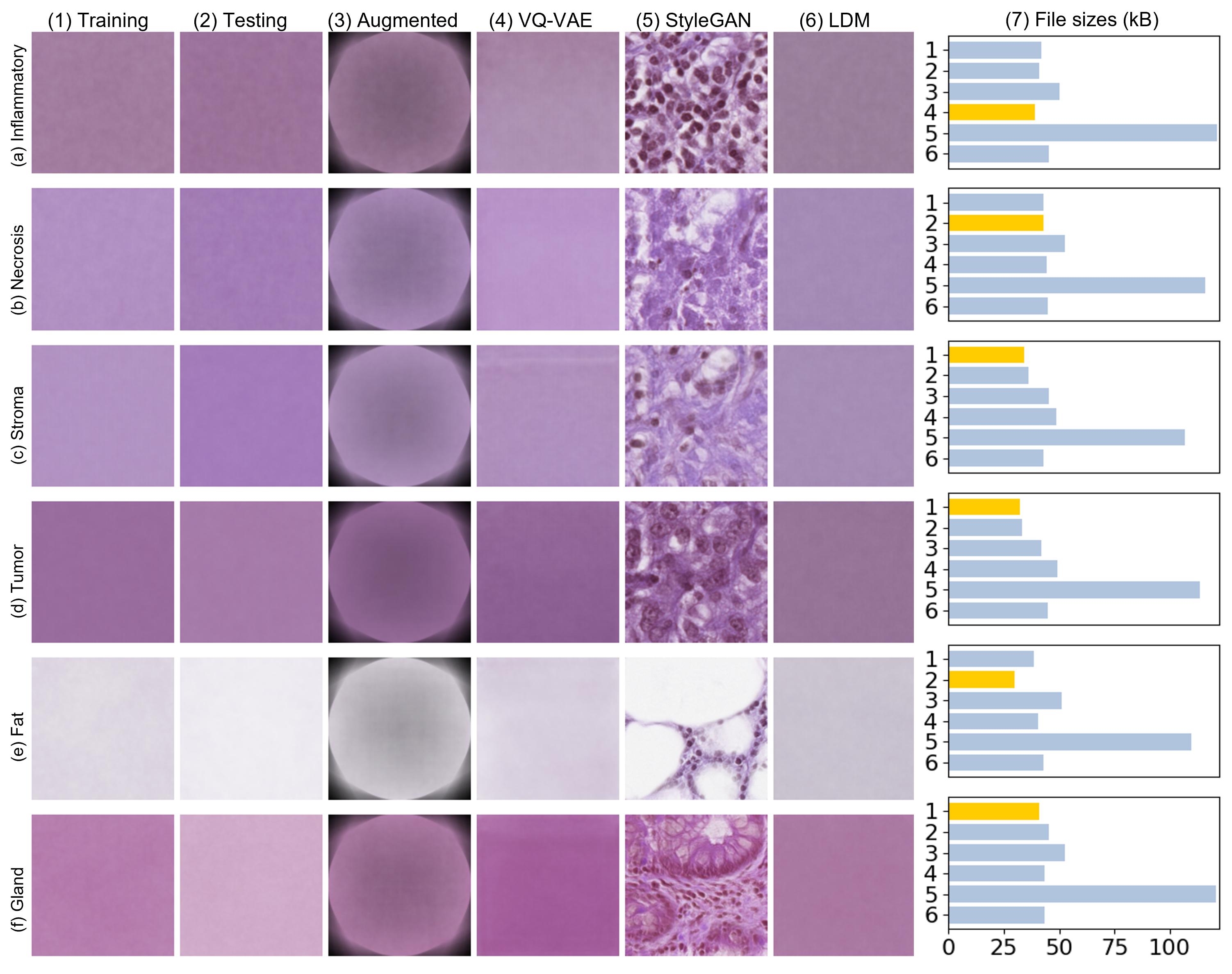}

\caption{Average images of synthetic and real datasets from the pathological dataset and their file sizes. The StyleGAN2 model had a mode collapse problem that produced images with similar structures.} \label{fig:pathology_avg}

\end{figure}

In comparison, GAN models produced synthetic images that could fool human experts the most: especially on the pathological dataset, only 1 out of 12 were successfully identified by human experts (0.08 FIR on Table \ref{tab:pathology}). However, a severe mode collapse problem was also found in the pathological dataset. Clear edges in the average images (Fig. \ref{fig:pathology_avg} (5)) indicate that the StyleGAN2 model produced images with the same spatial structures for each pathological category. As for the X-ray dataset, the mode collapse problem was not as severe as it was in the pathological dataset, while we could still observe a variety drop in the StyleGAN-synthesized images according to the increased file sizes of average image (from 50.46 to 115.62 in Table \ref{tab:xray}). The results demonstrated that the StyleGAN2 model could perform a higher quality data synthesis compared to other synthesis models on fully curated datasets \cite{sauer2022stylegan}, such as the well-aligned X-ray dataset in our work, while the performance dropped dramatically on unstructured datasets, such as the pathological dataset. 

Despite its high GPU memory costs, LDM could produce synthetic images with the best quality and utility on the pathological dataset. However, we noticed that the LDM failed to produce realistic images on greyscale images with high resolutions ($512\times512$) because of the high FIR (0.80 in Table \ref{tab:xray}). To further investigate the LDM performance, we trained LDMs on the additional CT dataset with different resolutions and the results are shown in Table \ref{tab:compare}. We discovered that LDM failed to produce good synthesis performance on the greyscale structured dataset, even though the resolution was as low as $256\times 256$. In comparison, StyleGAN2 achieved the best performance on the greyscale structured medical image datasets and had the capability to process high dimensional data such as $1024\times 1024$.

\begin{table}[h]
\caption{Fidelity and variety measurements of StyleGAN2 and LDM on greyscale structured datasets. We compared the image qualities between StyleGAN2 and LDM, and bold texts indicate a better quality compared to other methods in the same task setting.  }\label{tab:compare}
\resizebox{\linewidth}{!}{\begin{tabular}{|l|l|llllll|llll|}
\hline
\multirow{3}{*}{Dataset} & \multirow{3}{*}{Resolution} & \multicolumn{6}{l|}{Fidelity} & \multicolumn{4}{l|}{Variety} \\ \cline{3-12} 
 &  & \multicolumn{2}{l|}{FIR} & \multicolumn{2}{l|}{FID} & \multicolumn{2}{l|}{Precision} & \multicolumn{2}{l|}{Recall} & \multicolumn{2}{l|}{File size} \\ \cline{3-12} 
 &  & \multicolumn{1}{l|}{StyleGAN} & \multicolumn{1}{l|}{LDM} & \multicolumn{1}{l|}{StyleGAN} & \multicolumn{1}{l|}{LDM} & \multicolumn{1}{l|}{StyleGAN} & LDM & \multicolumn{1}{l|}{StyleGAN} & \multicolumn{1}{l|}{LDM} & \multicolumn{1}{l|}{StyleGAN} & LDM \\ \hline
\multirow{3}{*}{X-ray} & 256 & \multicolumn{1}{l|}{\textbf{0.50}} & \multicolumn{1}{l|}{0.80} & \multicolumn{1}{l|}{\textbf{31.97}} & \multicolumn{1}{l|}{79.71} & \multicolumn{1}{l|}{\textbf{0.79}} & 0.42 & \multicolumn{1}{l|}{\textbf{0.60}} & \multicolumn{1}{l|}{0.57} & \multicolumn{1}{l|}{34.83} & \textbf{23.28} \\ \cline{2-12} 
 & 512 & \multicolumn{1}{l|}{\textbf{0.50}} & \multicolumn{1}{l|}{0.80} & \multicolumn{1}{l|}{\textbf{9.85}} & \multicolumn{1}{l|}{25.68} & \multicolumn{1}{l|}{\textbf{0.68}} & 0.19 & \multicolumn{1}{l|}{\textbf{0.19}} & \multicolumn{1}{l|}{0.04} & \multicolumn{1}{l|}{92.84} & \textbf{70.59} \\ \cline{2-12} 
 & 1024 & \multicolumn{1}{l|}{\textbf{0.80}} & \multicolumn{1}{l|}{/} & \multicolumn{1}{l|}{\textbf{4.70}} & \multicolumn{1}{l|}{/} & \multicolumn{1}{l|}{\textbf{0.04}} & / & \multicolumn{1}{l|}{\textbf{0.00}} & \multicolumn{1}{l|}{/} & \multicolumn{1}{l|}{\textbf{94.4}} & \textbf{/} \\ \hline
\multirow{3}{*}{CT} & 256 & \multicolumn{1}{l|}{\textbf{0.25}} & \multicolumn{1}{l|}{1.00} & \multicolumn{1}{l|}{\textbf{41.76}} & \multicolumn{1}{l|}{51.69} & \multicolumn{1}{l|}{\textbf{0.40}} & 0.10 & \multicolumn{1}{l|}{0.00} & \multicolumn{1}{l|}{\textbf{0.19}} & \multicolumn{1}{l|}{30.52} & \textbf{21.96} \\ \cline{2-12} 
 & 512 & \multicolumn{1}{l|}{\textbf{0.25}} & \multicolumn{1}{l|}{1.00} & \multicolumn{1}{l|}{\textbf{8.24}} & \multicolumn{1}{l|}{14.16} & \multicolumn{1}{l|}{\textbf{0.20}} & 0.00 & \multicolumn{1}{l|}{0.00} & \multicolumn{1}{l|}{\textbf{0.02}} & \multicolumn{1}{l|}{95.35} & \textbf{47.38} \\ \cline{2-12} 
 & 1024 & \multicolumn{1}{l|}{\textbf{0.80}} & \multicolumn{1}{l|}{/} & \multicolumn{1}{l|}{\textbf{3.89}} & \multicolumn{1}{l|}{/} & \multicolumn{1}{l|}{\textbf{0.00}} & / & \multicolumn{1}{l|}{\textbf{0.00}} & \multicolumn{1}{l|}{\textbf{/}} & \multicolumn{1}{l|}{\textbf{97.15}} & \textbf{/} \\ \hline
\end{tabular}}

\end{table}
\begin{figure*}[t]

\includegraphics[width=\textwidth]{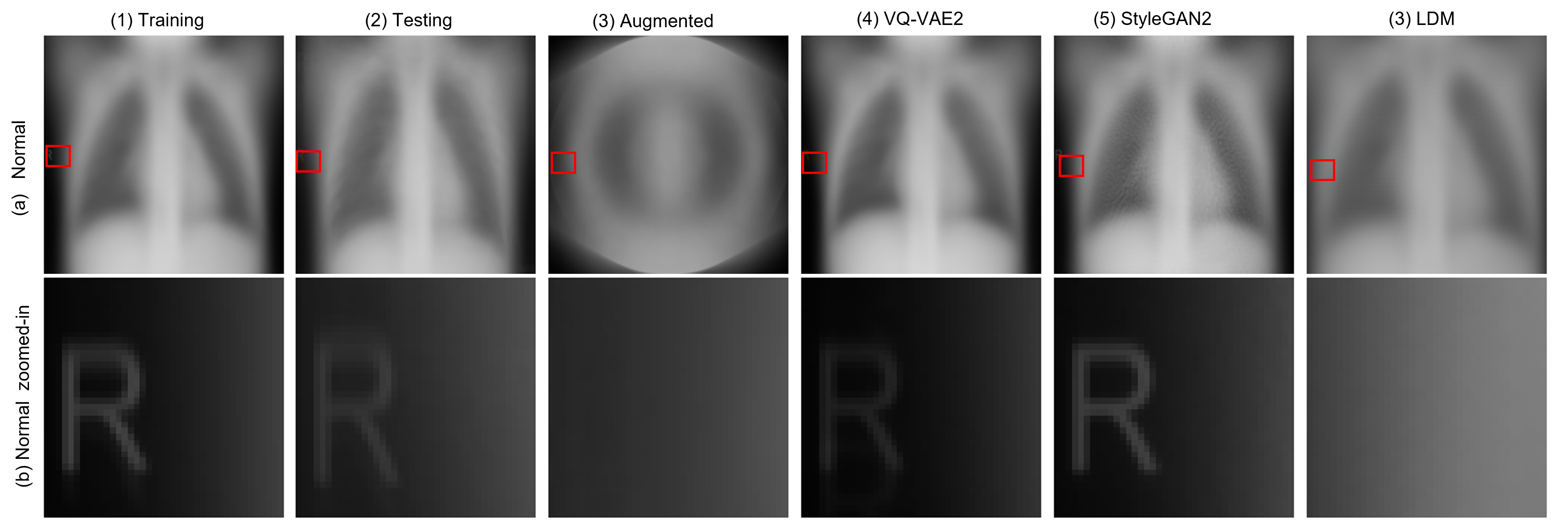}

\caption{The average images and zoomed-in images for highlighted regions for the X-ray dataset (Normal). The original dataset had a dataset bias problem where the locations of image texts were similiar on normal patients.} \label{fig:xray_avg}

\end{figure*}

\begin{figure*}[h]
\includegraphics[width=\textwidth]{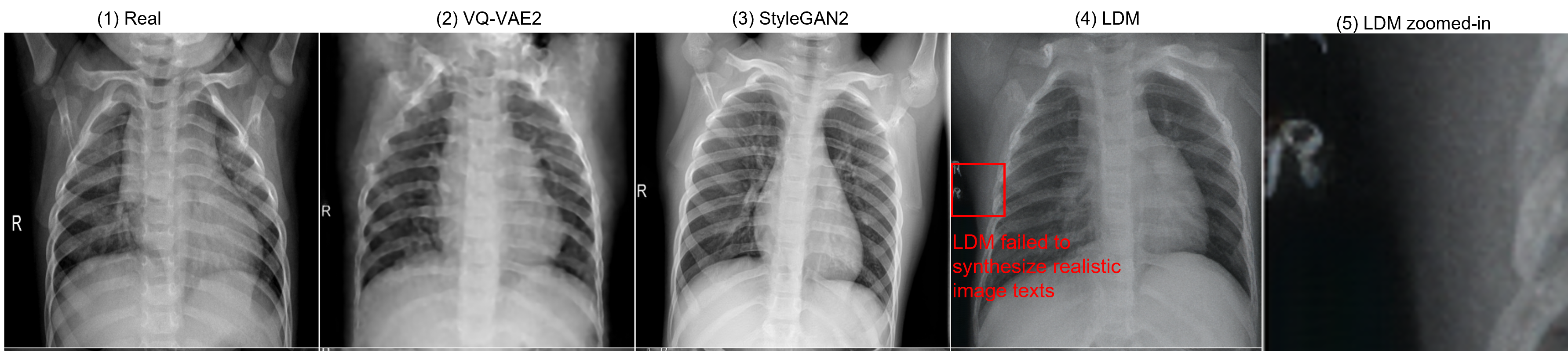}
\caption{Examples of X-ray datasets (Normal). The LDM model failed to synthesize realistic image texts, thus producing synthetic images that did not look real. } \label{fig:xray_example}
\end{figure*}

\subsection{Comparing Utilities of Different Synthesis Models}
From the utility measurement results in both Tables \ref{tab:pathology} and \ref{tab:xray}, we discovered that \textbf{on both datasets, the synthetic images failed to demonstrate a high data augmentation utility}. Only the synthetic pathological images from LDM could improve the classification accuracy. However, we discovered that the classification accuracy improvement brought by adding high-quality synthetic images (3.64\% in Table \ref{tab:xray}) has no significant difference ($p<0.05$) compared to traditional augmentations (4.07\% in Table \ref{tab:pathology}). Considering the high GPU memory cost and the training time cost, the value of using synthetic data for the data augmentation purpose was questionable. 

For the feature extraction utility, we discovered that \textbf{the feature extraction utility of synthetic images is limited} on the X-ray dataset because none of the synthetic images could train a feature extractor as accurately as the real images on the pathological dataset (Table \ref{tab:pathology}). 

We also discovered that \textbf{the utility is not correlated with fidelity}. The synthetic X-ray images from LDM were easy to identify, and 0.80 (Table \ref{tab:xray}) synthetic images were successfully identified by human experts. However, the LDM-synthesized X-ray can still train a feature extractor with compatible accuracy with the real images. 

\subsection{False Good Utility Score on StyleGAN2 Synthesized Images }

In Table \ref{tab:xray}, we noticed that synthetic images from the StyleGAN2 model could produce accurate feature extractor performance. However, this result did not indicate that the realistic images from the StyleGAN2 model had a high utility. As in Fig. \ref{fig:xray_avg}, we observed a dataset bias in the X-ray dataset. Interestingly, for normal patients, most of the image contained a letter R in similar regions, and models trained on this biased dataset tended to focus on the text regions, instead of actual lung lesions. Moreover, the GAN models inherited this dataset bias, reducing the robustness and reproducibility of the downstream tasks.

Fig. \ref{fig:xray_avg} shows that the LDM mode did not inherit the dataset bias. Thus, we visualized several example images from real and synthetic datasets. In Fig. \ref{fig:xray_avg}, the LDM model tried to capture the dataset bias, while it failed to synthesize realistic image texts and thus failed to capture the real but biased dataset distribution. This failure, however, turned out to increase the robustness of the feature extractor trained on the LDM-synthesized images.

To reveal the model faithfulness, we cropped the images to avoid the image texts as decisive factors, and the results are shown in Table \ref{tab:xraycropped}. After cropping out the image texts, the feature extraction utility of StyleGAN2 synthesized image dropped, while the LDM-synthesized images still obtained compatible feature extraction performance as the real dataset, indicating that the failure to capture the dataset bias increased the feature robustness of LDM synthesized images instead. In this example, we have shown that “realistic” images are not always useful. This result further validated our hypothesis that the utility is not correlated with fidelity because a faithful image generator might inherit the dataset bias, reducing the robustness and providing falsely good utility scores. \\
\begin{table}[h]
\caption{Fidelity, variety, utility, and efficiency measurements of all deep generative models in the cropped X-ray dataset (the image texts were cropped out). }\label{tab:xraycropped}
\resizebox{\linewidth}{!}{\begin{tabular}{|l|lll|ll|ll|ll|}
\hline
 & \multicolumn{3}{l|}{Fidelity} & \multicolumn{2}{l|}{Variety} & \multicolumn{2}{l|}{Utility} & \multicolumn{2}{l|}{Efficiency} \\ \cline{2-10} 
\multirow{-2}{*}{Method} &  \multicolumn{1}{l|}{FIR} & \multicolumn{1}{l|}{FID} & Precision & \multicolumn{1}{l|}{Recall} & File size & \multicolumn{1}{l|}{\begin{tabular}[c]{@{}l@{}}Augment \\ (\%)\end{tabular}} & \begin{tabular}[c]{@{}l@{}}Extraction\\  (\%)\end{tabular} & \multicolumn{1}{l|}{\begin{tabular}[c]{@{}l@{}}GB memory  \\ (8 images)\end{tabular}} & \begin{tabular}[c]{@{}l@{}}Inference time \\ (1000 images)\end{tabular} \\ \hline
Training & \multicolumn{1}{l|}{0.40} & \multicolumn{1}{l|}{9.07} & 0.73 & \multicolumn{1}{l|}{0.84} & 38.4 & \multicolumn{1}{l|}{83.49} & 83.49 & \multicolumn{1}{l|}{/} & / \\ \hline
Augmented & \multicolumn{1}{l|}{{\color[HTML]{FF0000} 0.50}} & \multicolumn{1}{l|}{{\color[HTML]{FF0000} 185.89}} & {\color[HTML]{FF0000} 0.04} & \multicolumn{1}{l|}{0.71} & 36.89 & \multicolumn{1}{l|}{(+) 6.41} & / & \multicolumn{1}{l|}{/} & / \\ \hline
VQ-VAE2 & \multicolumn{1}{l|}{{\color[HTML]{FF0000} 1.00}} & \multicolumn{1}{l|}{{\color[HTML]{FF0000} 81.59}} & {\color[HTML]{FF0000} 0.07} & \multicolumn{1}{l|}{{\color[HTML]{FFC000} 0.00}} & 41.23 & \multicolumn{1}{l|}{{\color[HTML]{00B050} (+) 1.60}} & {\color[HTML]{00B050} (-) 20.99*} & \multicolumn{1}{l|}{{\color[HTML]{0070C0} 22.82}} & {\color[HTML]{0070C0} 12h 30min} \\ \hline
StyleGAN2 & \multicolumn{1}{l|}{{\color[HTML]{FF0000} 70.00}} & \multicolumn{1}{l|}{14.08} & 0.77 & \multicolumn{1}{l|}{{\color[HTML]{FFC000} 0.31}} & {\color[HTML]{FFC000} 92.8} & \multicolumn{1}{l|}{{\color[HTML]{00B050} (+) 1.44}} & {\color[HTML]{00B050} (-)13.30*} & \multicolumn{1}{l|}{1.19} & 1 min \\ \hline
LDM & \multicolumn{1}{l|}{{\color[HTML]{FF0000} 0.80}} & \multicolumn{1}{l|}{{\color[HTML]{FF0000} 39.08}} & {\color[HTML]{FF0000} 0.17} & \multicolumn{1}{l|}{{\color[HTML]{FFC000} 0.01}} & {\color[HTML]{FFC000} 52.81} & \multicolumn{1}{l|}{(+) 8.01*} & (-) 0.64 & \multicolumn{1}{l|}{{\color[HTML]{0070C0} 70.61}} & {\color[HTML]{0070C0} 39 min} \\ \hline
\end{tabular}}

\end{table}
\textcolor{white}{a}\\
\textcolor{white}{a}\\

\section{Conclusion}
In this study, we conducted an empirical evaluation of three major types of deep generative models and measured the correlation between synthetic image quality and utility. Our analysis revealed that diffusion models failed to generate realistic images for the X-ray dataset containing structured greyscale images. Additionally, diffusion models required a large amount of GPU resources for training and optimization, and their inference time increased dramatically with image resolution. While our work demonstrated a high utility of diffusion model synthesized medical images, the lack of fidelity and high computational costs raise questions about their real-world application in medical image synthesis for data augmentation in high-resolution medical image classification tasks.

Furthermore, we found that metrics used to evaluate the quality of synthetic images were questionable, as images with high-quality scores may have low utility for downstream tasks. In contrast, synthetic images can produce high classification accuracy even if they lack realism or variety. Our study demonstrates that synthetic image utility cannot be measured without performing downstream tasks. Rather than blindly using synthesized images from deep generative models, we propose the development of utility-aware and explainable models for medical image synthesis. These models can help address the shortcomings of current deep generative models and improve the utility and applicability of synthetic images for medical image classification tasks.

\subsubsection{Acknowledgements} This study was supported in part by the ERC IMI (101005122), the H2020 (952172), the MRC (MC/PC/21013), the Royal Society (IEC/NSFC/211235), the NVIDIA Academic Hardware Grant Program, the SABER project supported by Boehringer Ingelheim Ltd, NIHR Imperial Biomedical Research Centre (RDA01), and the UKRI Future Leaders Fellowship (MR/V023799/1).

%
%
%
\bibliographystyle{splncs04}
\bibliography{ref}

\end{document}